\documentclass[12pt,onecolumn,showpacs,amssymb,aps,nofootinbib,floatfix]{revtex4-1}
\usepackage{epsfig}
\newcommand{\ave}[1]{\left\langle #1 \right\rangle}

\newcommand{\eqcomma}{\phantom{AA},\phantom{AA}}

\newcommand{\modulo}[1]{\left| #1 \right|}
\newcommand{\order}[1]{ \mathcal{O} \left( #1 \right) }

\begin{document}
\title{Holography in a background-independent effective theory}
\author{Giorgio Torrieri}
\affiliation{Universidade Estadual de Campinas - Instituto de Fisica "Gleb Wataghin"\\
Rua Sérgio Buarque de Holanda, 777\\
 CEP 13083-859 - Campinas SP\\
torrieri@phys.columbia.edu\\
%Essay written for the Gravity Research Foundation 2014 Awards for Essays on Gravitation\\
%Date of submission:  2/26/2014.   Country of submission: Brazil
}
\begin{abstract}
We discuss the meaning of the strong equivalence principle when applied to a quantum field theory.
We show that, because of unitary inequivalence of accelerated frames, the only way for the strong equivalence principle to apply exactly is to add a boundary term representing the decoherence of degrees of freedom leaving the observable region of the bulk.
We formulate the constraints necessary for the partition function to be covariant with respect to non-inertial transformations and argue that, when the non-unitary part is expressed as a functional integral over the horizon, holography arises naturally as a consequence of the equivalence principle. 
\end{abstract}
\maketitle
\section{Introduction: the equivalence principle and quantum field theory}
The central assumption of general relativity is the strong equivalence principle (henceforward referred to equivalence principle or EP)\cite{ep}:  a freely falling frame is indistinguishable locally from an inertial frame with a Minkowski metric.   Viceversa, stationary frames in gravitational fields are indistinguishable from accelerating frames, represented in Minkowski space by coordinate systems with curvature.
The strong EP implies that {\em all} laws of physics respect this invariance.   
This implies that all laws of physics have to be written covariantly with respect to non-inertial transformations, and that gravitational fields have to be described in terms of a curved spacetime, in which {\em all} metrics are warped. 

While the EP leads to an elegant picture of {\em classical} physical laws, 
 extending it to the structure of quantum field theory, which reconciled quantum mechanics\footnote{For the purposes of this introduction, ``quantum mechanics'' refers to the description of states $|\psi>$ in terms of vectors in Hilbert space evolving linearly under the action of unitary operators, $|\psi> \rightarrow \exp[i\hat{A}] |\psi>$, and observables as hermitian operators ($\sim \hat{A}$) defined on this space.  ``Quantum field theory'' refers to the extension of this framework to fields, valued over each point in spacetime.  As we argue toward the end of the introduction, this definition is reductive since hermiticity, linearity and unitarity are only applicable for {\em closed} quantum systems.  However, a rigorous definition of quantum mechanics without this constraint is currently lacking (the reformulation in terms of sum-over-paths and partition functions comes closest to to such a rigorous definition), so weather the ``modification of quantum mechanics'' discussed this work is really a modification of quantum mechanics is a matter of opinion and semantics   } with {\em special} relativity  \cite{qft,weinberg3} has proven very difficult, with both technical issues, such as non-renormalizeability of gravity as a field theory \cite{weinberg3}, and conceptual paradoxes, such as the black hole information problem \cite{blackinfo}, being highly controversial after decades of study.

This should not surprise us:
 The mere fact that \cite{unruh,hawking,fulling,grove}, a detector in curved space, whether due to acceleration (usually called the Unruh effect) or a gravitational field (usually called Hawking radiation), sees a finite temperature bath makes combining background independence with quantum mechanics fundamentally problematic, because even infinitesimal non-inertial deformations put the
vacuum in another superselection sector with a different entropy content.
Note that this signals an ambiguity of the semiclassical limit of QFT as well as gravity:
 Haag's theorem \cite{haag} ensures that even an infinitesimal coordinate deformation \cite{biso,sewell} will send the vacuum of the theory into a unitarily inequivalent orthogonal state \cite{fulling,grove}, one which is indistinguishable from a thermal mixed state which cannot be reached by quantum evolution.  For normal QFT this ambiguity is irrelevant as this unitary inequivalence can be reabsorbed into field-strength renormalization, but for a theory of gravity such a renormalization would inevitably modify observable asymptotic states
\footnote{The reasoning here is similar to the proof that an infinitesimal deformation of the Hamiltonian produces orthogonal eigenstates \cite{haag}.   However, while for hamiltonian deformations such infinities are known to be regularizeable \cite{qft} under certain conditions ( notably, the fact that the possible deformations respect the underlying spacetime and internal symmetries.) No similar regularizing procedure can exist for non-inertial coordinate deformations, as these deformations typically change asymptotic states and make the LSZ \cite{qft} reduction formula inapplicable.
The fact that the deformation appears thermal was proved in the context of axiomatic field theory in \cite{biso,sewell}
}.     The universal nature of the ``fragility'' of quantum field theory to deformations is illustrated by the fact that interactions and the Unruh effect are deeply connected: {\em dynamically} the Unruh effect can be understood, in an inertial frame, in terms of QFT interactions (Bremstrahlung-like processes to leading order) between a charged detector and the field provoking its acceleration  \cite{higuchi1,higuchi2,sudarsky}.  
In this picture the different entropy content of the system can be understood in terms of the use of the semiclassical approximation and decoherence \cite{boulware,saa} with respect to the zero entropy Minkowski vacuum.

\begin{figure}[h]
\epsfig{width=0.84\textwidth,figure=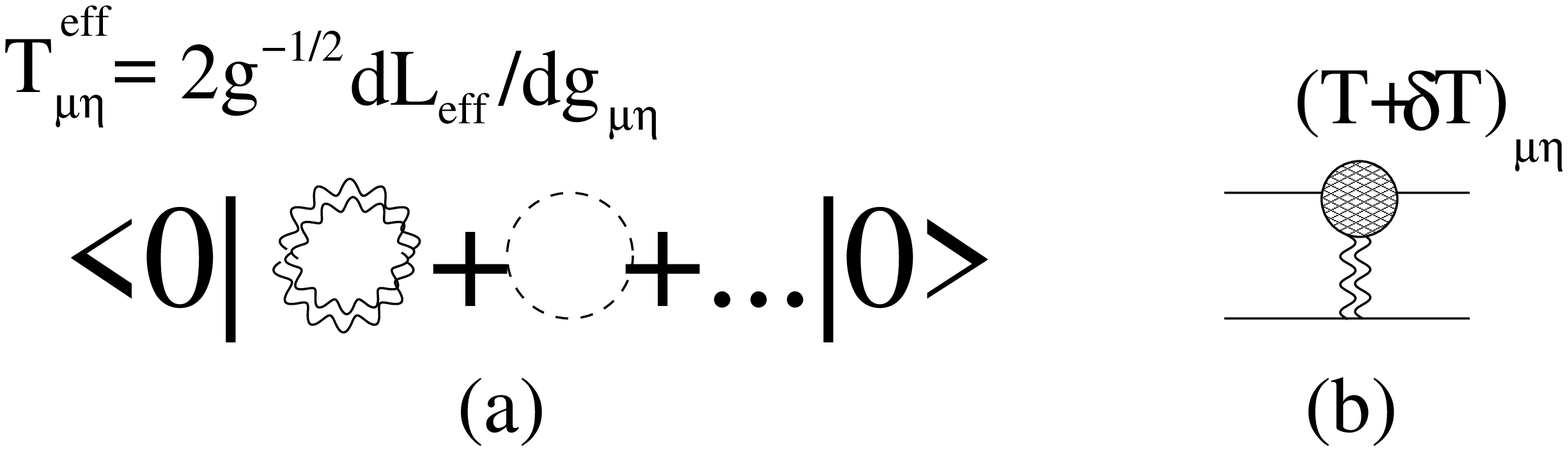}
\caption{\label{def} The two definitions of the energy momentum tensor, in terms of the effective Lagrangian including vacuum corrections (panel (a)) and the graviton vertex function (panel (b)).   The strong equivalence principle requires these to be {\em strictly} equal, something generally not true for gravity with quantum corrections (see for example \cite{soph})}
\end{figure}

This means, however, that an ``accelerated'' transformation will send the system into a different
vacuum, a non-unitary transformation, as made evident by the different entropy content of this vacuum.   A dynamical change in gravitational fields,however, needs to be unitary if gravity is to be a quantum theory as usually defined (see footnote 1).   While a linearized gravitational fluctuation maintains EP (in fact, is required by quantum consistency to do so \cite{weinberg3}), any vacuum backreaction should break it.
Quantitatively, this should manifest itself as an anomalous difference between the energy-momentum tensor defined from the effective action $T_{\mu \nu}^{eff}$ and the vertex function for the graviton $T_{\mu \nu} +\delta T_{\mu \nu}$ (Fig. \ref{def} panels (a) and (b) respectively).
%%%%%%%%%%%%%%%%%%%%%%%%%%%%%%
%%%%%%%%%%%%%%%%%%%%%%%%%%%%%%%%%%%%%%%%%%%%%%%
This, in fact,has long been seen explicitly in loop corrections of spin-2 theories \cite{soph}\footnote{The latest such calculation is \cite{soph}, where, after a tour-de-force computation it was found that light deflection by gravity depends on spin at one loop.   In the discussion of that paper it is claimed that in this regime such a measurement becomes a non-local because the Compton wavelength becomes comparable to the curvature.  However, as these are point particles, this is another way of saying that in this limit local measurements are impossible, i.e., no physically relevant equivalence principle can be defined }, and even at the level for first quantization \cite{das}.
Note that lack of unitarity has nothing to do with energy conservation ( in an accelerated frame the Hamiltonian has an explicit time dependence), but is dependent on the fact that quantum field theories are local and have infinitely many degrees of freedom.  In a fundamental theory where degrees of freedom at the Planck scale are countable, non-inertial transformations should also be unitary, and it is not clear whether the Unruh effect persists in such models \cite{nounruh1,nounruh2}.   
If it does not, however, the existence within such theories of a semiclassical limit of a quantum field theory living on a manifold is doubtful because of the connection of the Unruh effect to QFT in Minkowski space \cite{higuchi1,higuchi2,sudarsky}.

Let us try to interpret what this incompatibility means in a more physical way.
In classical physics, while non-inertial frames are distinguishable from inertial ones, dynamically the difference can be parametrized by fictitious forces (Coriolis, centrifugal and others, represented by parallel transport formalisms: Christoffel symbols and so on).  The EP implies that
\begin{itemize}
\item Any non-gravitational consequences of choosing a non-inertial frame can be parametrized by such a formalism centered around parallel transport
\item Choosing a non-inertial frame in and of itself cannot impact the gravitational dynamics, which is represented by the geometry of space and not the local metric choice.  
\end{itemize}
If this is to be a semiclassical limit of a quantum theory, either the role of the horizon implies a general breaking of unitarity or the gravitational forces cannot be fully encoded this way:
In Minkowski space, and {\em only} in Minkowski space, any correlations at superhorizon distances will cancel out when antiparticles and particles are taken into account \cite{qft} and so we can say that the bulk and the horizon are fully decoupled.
The violation of unitarity implicit in the Unruh effect can be interpreted \cite{boulware,saa} as the tracing over of degrees of freedom lying beyond a Rindler horizon.  If this evolution is to be unitary, this tracing over can only be an approximation, and hence an observer could in principle reconstruct states appearing beyond the horizon exactly (This reasoning applied to a black hole geometry is the essential issue behind the so-called black-hole information paradox \cite{blackinfo}).  By definition such a horizon is not a ``true'' horizon but an apparent one, which ceases being a horizon when the right observable is measured.  Since horizons encode information on the curvature of geometry \cite{ray,jacobson,verlinde} (including infinitesimal perturbations \cite{bianchipaper,wald}), and the strong EP implies gravity to be geometric, the EP requires {\em no} observable can be constructed sensitive to degrees of freedom beyond the horizon (the horizon must indeed be a ``true'' one) and hence the tracing over needs to be {\em exact}. 
this suggests {\em either} the equivalence principle, {\em or}  quantum 
mechanics {\em or} at least observer-specific unitarity have to be abandoned.   

Aesthetic and consistency grounds make for skepticism of the second alternative, which nevertheless has some following (see e.g. \cite{snewton});   
Most of the practitioners of the field tend to prefer the first alternative, sometimes couching it in arguments such as those in \cite{soph}.  Indeed, high precision empirical tests of the strong equivalence principle are not to at the level sufficient to test this alternative, which would require precision measurements of the difference highlighted in Fig. \ref{def} (Precise torsion experiments \cite{review,tors} tend to test the weak version. To the author's knowledge the only direct tests that  self-energy respects the EP are Nordveldt effect measurements \cite{nor1,nor2}).      Given the historical role of the {\em exact} EP in the development of general relativity as geometric theory, however, it is tempting to discuss ``how much of quantum mechanics'' one has to abandon for the EP to hold {\em at all orders}, in $\hbar$ and field coupling constants if not $l_p$.
\subsection{ An EP-invariant quantum field theory of {\em open} systems}
To impose the strong equivalence principle means making sure the two local detectors, instantaneously in two different (not necessarily inertial) coordinate systems will obtain the same values for {\em all observables} once the non-inertial effects (the equivalent of the pseudo-frces in the classical description) are ``properly'' accounted for.   The crucial issue is to understand the meaning of this ``properly'' within a quantum field theory:  
In the sum-over-paths picture, it means adding the imaginary {\em boundary term} to the action \cite{gibbons}, so what appears as a unitary quantum evolution to one observer will appear as a partially or fully decohered ``thermal''
 evolution to the other.

   This means that covariance with respect to non inertial transformations implies it should be possible to substitute a ``thermal'' ensemble for a quantum one by going to a different coordinate system.     This is {\em not} incompatible with standard quantum mechanics: Horizons make any quantum system {\em open} (the degrees of freedom outside the horizon have to be traced over), and open quantum systems have been studied for decades \cite{feynman,zurek,caldeira}. Objects such as partition functions, correlators and density matrices, remain well-defined.  
Since curvature of space is necessarily linked to the appearance of a causally connected horizon \cite{ray,jacobson,verlinde}, this is a promising approach for encoding the equivalence principle at the level of an Effective Field Theory (EFT).  
As we will show, holography \cite{holo,holo1,holo2} follows naturally as a consequence. While we are not aware of such an EFT approach being proposed before, a similar argument was made in \cite{padma,padma2} at the {\em classical} level. Furthermore, given the definition of local geometry in terms of the horizon \cite{ray,jacobson},  our approach is compatible with the thermal time
hypothesis \cite{rovelli2}/the Wootters-Page decoherence time
picture \cite{wp1,wp2,wp3,wp4} \footnote{
This connection, previously noticed in \cite{chapline} is central to a crucial issue of compatibility of quantum mechanics with general relativity, the question of time:  In General Relativity, the time direction is to a certain extent a Gauge choice, which can be changed by a non-inertial transformation.   In {\em any} probabilistic theory where the amount of information in possession of the observer changes (the Monty Hall problem, random walks etc) information defines an objective time direction.  In Quantum Mechanics this is implemented via unitarity, hence the connection to \cite{rovelli2,wp1,wp2,wp3,wp4}, as long as the generalized second law of
thermodynamics, hinted at from particular solutions of Einstein's equations
\cite{bekenstein,casini,davies1,davies2},  holds for arbitrary backgrounds}, and the mathematical formalism is similar to that described in section IIIC of \cite{ashtekar} and references therein. In addition, there is a deep mathematical analogy between the physics advocated here and the Gribov horizon picture in nonabelian gauge theores \cite{gribov}.
\section{EP-derived Constraints from the sum-over-paths picture}
We shall proceed in the Feynman sum-over-paths picture \cite{qft} and define our EFT at the level of the partition function.
Given a d-dimensional bulk metric $g_{\Sigma}^{\mu \nu}$ with local light-cone variables, a horizon, defined as $x^\mu \in \Sigma, \forall x^\mu, g^{\mu \nu}_\Sigma dx_\mu dx_\nu =0$, can be viewed as a d-1 dimensional surface $\partial \Sigma$  with coordinates $\sigma^i$ and local metric $g^{i j}_{\partial \Sigma}$ \cite{bianchipaper} (bulk coordinates are in Greek, surface coordinates in Latin).  
In particular, given \cite{bianchipaper} a d-dimensional bulk metric $g_{\Sigma}^{\mu \nu}$ with local light-cone variables $z_{+,-}=x \pm t$, the horizon, defined as $x^\mu \in \Sigma, \forall x^\mu, g^{\mu \nu} dx_\mu dx_\nu =0$, can be viewed as a d-1 dimensional surface $\partial \Sigma$ with coordinates $\sigma^i$ and local metric  $g^{i
j}_{\partial \Sigma}$ related by
\begin{equation}
\label{horizondef}
g_{\partial \Sigma}^{i j} = \gamma_{\mu \nu} \left( \frac{\partial x^\mu}{\partial \sigma^i}\frac{\partial x^\nu}{\partial \sigma^j} - \frac{\partial
x^\mu}{\partial \sigma^j}\frac{\partial x^\nu}{\partial \sigma^i}\right) \eqcomma \gamma_{\mu \nu} = g_{\mu \nu} + \frac{1}{2}\left( \frac{\partial x_\mu}{\partial
z_+} \frac{\partial x_\nu}{\partial z_{-}} + \frac{\partial x_\mu}{\partial z_{-}} \frac{\partial x_\nu}{\partial z_+} \right)
\end{equation}
beyond Eq. \ref{horizondef},
there is no general procedure for constructing $\sigma_i,g_{\partial \Sigma}^{ij}$ given a $g^{\mu \nu}_\Sigma$ and,
 for most geometries, $g_{\partial \Sigma}^{ij}$ has very little in common with the Minkowski metric on which realistic quantum field theories is defined, since this is a degenerate metric admitting no comoving time direction \cite{kuchar,nondeg0,nondeg01} (By diagonalizing the non-zero components, one can reduce it to a lower-dimensional euclidean metric), and generally not admitting any global symmetries \cite{kuchar}. The sum-over-paths picture, however, allows us to define Quantum field theories on such manifolds in a relatively straight-forward manner \cite{euclidean}.
As one can always define a horizon from a near-horizon geometry (see for instance \cite{nondeg} section 2), one can conceptually construct the ``partition function of the universe as observed by a detector'' moving in a d-dimensional space with a {\em not} necessarily Minkowski metric $g^{\mu \nu}_{\Sigma}$.    

Let us assume the semiclassical limit, where the time it takes for a DoF to cross the horizon and decohere  is ``short'' ($\sim l_p$, the Planck length, in unites where $\hbar=c=1$) with respect to the evolution of the metric. 
One can then impose constrains that ``all detectors'' at the same point in spacetime in non-inertial reference frames will measure the same correlation functions.   

The action in a certain reference frame, will have a quantum bulk and decohered horizon component
\begin{equation}
S\left( g_\Sigma^{\mu \nu}(x),g_{\partial \Sigma}^{i j}(\sigma),\phi_\Sigma(x),\phi_{\partial \Sigma} (\sigma) \right) =  \underbrace{S_{\Sigma}(\phi_\Sigma)}_{\sim \order{l_p^0}} + \underbrace{i S_{\partial \Sigma}(\phi_{\partial \Sigma})}_{\sim \order{l_p^{2}}} + \underbrace{S_{int}}_{\sim \order{l_p^{n>2}}}(\phi_\Sigma,\phi_{\partial \Sigma})
\label{sdef}
\end{equation}
In terms of the respective fields, $\phi_\Sigma$ and $\phi_{\partial \Sigma}$
\begin{equation}
\label{bulks}
S_{\Sigma}= \int_\Sigma \sqrt{g_\Sigma} d^n x  L_{\Sigma} \left(\phi_\Sigma (x_\mu ) \right)
\eqcomma S_{\partial \Sigma } = \frac{1}{T} \int_{\partial \Sigma} F\left( \phi_{\partial \Sigma} (\sigma) \right) \sqrt{g_{\partial \Sigma}} d^{n-1} \sigma_i   
\end{equation}
 $T$ is the horizon temperature, determined through the usual area law
 \cite{jacobson} in terms of the entropy $\mathcal{S}_{\partial \Sigma}$.
\begin{equation}
\label{horizonT}
T =  \frac{dE_{\Sigma}}{d\mathcal{S}_{\partial \Sigma}} =- \frac{dE_{\partial \Sigma}}{d\mathcal{S}_{\partial \Sigma}}
\end{equation}
where the energy flow across the horizon is readable from Eq \ref{bulks} in terms of the bulk and boundary stress-energy tensors $T_{\Sigma}^{\mu \nu},T_{\partial \Sigma}^{i,j}$ constructed from $L_\Sigma,F_{\partial \Sigma}$
\begin{equation}
\label{stresse}
  E_{\Sigma} = \int_{\Sigma} T^{\mu \nu}_\Sigma \frac{\partial d x_\mu \partial d x_\nu}{\partial d \sigma_i d \sigma_j} d\sigma^i d \sigma^j \eqcomma E_{\partial \Sigma} = \int_{\partial \Sigma} T_{i j}^{\partial \Sigma}\sqrt{g_{\partial \Sigma}} d\sigma^i d\sigma^j
\end{equation}
\begin{equation}
 \mathcal{S}_{\partial \Sigma} =l_p^2 \int_{\partial \Sigma} \sqrt{g_{\Sigma}^{ij}} \prod d\sigma_i
\end{equation}
thus, $T$ serves as a Lagrange multiplier ensuring approximate local equilibrium between bulk and boundary.
Note that in Minkowski space and around a black hole, respectively, the real and imaginary parts $S_\Sigma$ and $S_{\partial \Sigma}$ become negligible.  

I the semiclassical limit in Eq. \ref{sdef} we have $\modulo{S_{int}}\ll \modulo{S_{\partial \Sigma}}, \modulo{S_\Sigma}$ so we could neglect $S_{int}$ and $S_{\Sigma},S_{\partial \Sigma}$ decouple (analogously to two cells in a perfect fluid, they are in instantaneous equilibrium with respect to each other and thus the dynamics of each appears as independent of the other.  Note that ``instantaneous'' refers to the detectors comoving frame)\footnote{ 
Note that in conventional holography, in the planar limit, this term is negligible, as the horizon is unaffected by Hawking radiation emanating from the bulk and vice-versa.     While this limit corresponds naturally to classical gravity (see discussion around Eq. \ref{heisen}), 
for global entropy estimates, suppressed terms most likely are relevant \cite{gteft,gfluid}}.

The necessity of the surface term $S_{\partial \Sigma}$ \cite{gibbons} and the temperature \cite{unruh,hawking} have long been known, but the discussion here gives them an interpretation rooted in the equivalence principle.
The equivalence principle for the {\em quantum} theory implies the covariance of all correlation functions with respect to non-inertial transformations.   In both unitary and finite temperature (fully decohered) quantum mechanics such correlations can be represented as functional derivatives of a partition function $Z$, expressed as a functional integral over all possible field convigurations and a scalar under non-inertial transformations\cite{qft}
\[\    
\left(\ave{\phi(x_1)...\phi(x_n)} \sim \frac{\delta^n \ln Z}{\delta J^{n}_{\Sigma,\partial \Sigma}},\mathrm{where\phantom{A}J\phantom{A}is\phantom{A}a\phantom{A}source\phantom{A}term}\right)\]    
In the semiclassical limit where $S_{int}$ is negligible, $Z$ will have to be defined both in the bulk and the horizon, each with its source term, as
\begin{equation}
\label{zdef}
Z =\int \mathcal{D} \phi_\Sigma \mathcal{D} \phi_{\partial \Sigma} \exp \left[\int \sqrt{g}_\Sigma d^4 x \left(i L + J_{\Sigma}(x)\phi_\Sigma \right) - \frac{1}{T} \int \sqrt{g}_{\partial \Sigma} d^3 \sigma \left(  F(\phi_{\partial \Sigma})+ J_{\partial \Sigma}(\sigma) \phi_{\partial \Sigma}(\sigma) \right) \right]
\end{equation}
Under arbitrary coordinate transformations, $x^\mu \rightarrow x'^\nu, g'_{\alpha \beta} = (dx'_\alpha /dx^\nu)  (dx'_\beta /dx^\nu) g^{\mu \nu}$, with a corresponding shift in $\sigma_i$ (see Eq. \ref{horizondef}).   For $Z$ to be invariant under such transformations it is required that at each point $x$
\begin{equation}
\label{constraint1}
\delta_{x^\mu \rightarrow  x'^\nu} Z\left( \Sigma,\partial_\Sigma,J_\Sigma (x),J_{\partial \Sigma}(\sigma),T\right) = \mathcal{K}_0 Z\left( \Sigma',\partial \Sigma',J_\Sigma (x'),J_{\partial \Sigma}(\sigma'),T'\right)
\end{equation}
where $\mathcal{K}_0$ is at most a constant, possibly divergent rescaling which factors out of the correlation functions.

Note that Equation \ref{sdef} implies that in the classical limit only bulk-bulk and boundary-boundary correlators will be relevant, but beyond it bulk-boundary correlations, controlled by $S_{int}$, will be possible.
\section{Holography from background independence}
Equation \ref{constraint1} ensures the {\em invariance } of the partition function with respect to non-inertial transformations (Eq. \ref{zdef} ensures the {\em covariance})\footnote{We note that holographic setups studied in the literature are covariant 
but not invariant under diffeomorphism transformations, since the horizon 
relevant for the correspondence is defined in a preferred inertial frame}.
 Together with a semiclassical geometric largrangian ($L_{gravity} \sim f(R)$), the equilibrium condition Eq. \ref{horizonT} and the boundary definition of  Eq. \ref{horizondef} it can provide a definition of a theory obeying the equivalence principle at the quantum level:   A transformation will shift ``action'' to and from the bulk theory living on $\Sigma$ to a boundary theory living on $\partial \Sigma$, and change the temperature $T$.
Just like classical covariance implies the presence of objects such as coordinate-depndent affine connections in all dynamical equations of motion, quantum covariance implies the presence of coordinate-depndent horizons in partition functions.
     Dualities such as \cite{holo}, and more generally the holographic principle \cite{holo1,holo2}, arise naturally  in the EFT given two coordinate systems $x_{1,2}$ in which $S_{int}$ is negligible.    The first system is bulk-dominated, the second boundary-dominated, as in this limit Equation \ref{constraint1} implies
\begin{equation}
\label{constraint3}
\int \mathcal{D} \phi_{\Sigma_1} \exp \left[  \int \sqrt{g_{\Sigma_1}} d^n x \left(  L_{\Sigma_1} \left(\phi_{\Sigma_1}  \right) + J_{\Sigma_1(x)} \phi_{\Sigma_1(x)} \right) \right] \simeq
\end{equation}
\[\
 \mathcal{K}_0   \int\mathcal{D} \phi_{\partial \Sigma_2}  \exp \left[-\frac{1}{T} \int_{\partial \Sigma_2}  \sqrt{g_{\partial \Sigma_2}} d^{n-1} \sigma \left(  F(\phi_{\partial \Sigma_2}) +J_{\partial \Sigma_2}(\sigma ) \phi_{\partial \Sigma_2}  \right) \right]
\]
In Minkowski space inertial frames will have a completely decoupled horizon, since it is causally inaccessible due to the presence of antiparticles \cite{qft} (
 any acceleration will however induce a rindler horizon which will act as a boundary). In curved spacetimes, where the Minkowski frame can be defined only locally, however, two such frames are indeed possible:   Consider the stationary frame and the freely-falling Kruskal frame in a black hole \cite{wald}.   In the first, the dominant component of the entropy is the horizon, in the second case it is the bulk.  

A consequence of Eq. \ref{constraint3} would be a ``dictionary'', linking bulk operators, defined as functional derivatives of $\frac{\delta^n \ln Z(L(\phi_{\Sigma_1}))}{\delta \phi_{\Sigma_1}^n}$, with surface operators defined as functional derivatives of  $\frac{\delta^n \ln Z(L(\phi_{\partial \Sigma_1}))}{\delta \phi_{\partial \Sigma_1}^n}$.
For both of these theories to be ``physical'' quantum field theories, both $\Sigma$ and $\partial \Sigma$ have to be time-like.   In physical space this should {\em not} occur because of the horizon's absence of global symmetries \cite{kuchar}.   However, this is not a fundamental obstacle to the mathematical definition of the dual theory since the sum-over-paths picture allows us to define a theory without a time-coordinate \cite{euclidean}, and even apply it to a holographic setting \cite{xiao}.  The correlators at the bulk and the boundary will just be related by a complicated, and metric-dependent, coordinate transformation involving both space and time;  In particular, for spacetimes with a null, or positive cosmological constant, the degeneracy of the metric will mean that the dual theory can be seen as a euclidean, generally highly asymmetric theory in $D-2$ dimensions.
In a space, such as AdS space, where the boundary is specifically time-like, however, equation \ref{constraint3} yields two time-like theories, defined in $D+1$ and $D$-dimensional space.  This is exactly what one would expect from the way holography is usually defined \cite{holo}, and makes it possible for $F(\phi_{\partial \Sigma_2})$ and $L_{\Sigma_1} \left(\phi_{\Sigma_1}  \right)$ to be separately unitary (Equivalently, the free limits of bulk and boundary theories are isomorphic \cite{holo}).    In this scenario, the conformal symmetry of the horizon constrains the theory there to a CFT, while, if unitarity is imposed, the background independence of the bulk is reduced, for those frames where unitary holography is valid, to the invariance w.r.t. diffeomorphisms that leave the asymptotic behavior invariant.

If, in addition, one of the two theories is weakly coupled, the holography in question becomes computationally useful.
 If the conjectures described here apply, however, such setups are a leading order approximation of highly symmetric setups, within a more general description including {\em both} the bulk and the boundary, rather than the popular conjecture that ``the universe is a hologram'' describable in its entirety by a lower dimensional theory \cite{holo1,holo2}.

It should be noted that, due to the Weinberg-Witten theorem, a self-consistent spin-2 theory will exhibit the equivalence principle at tree level \cite{weinberg3}\footnote{This can be seen by the curvature kinetic energy and the coupling of the graviton to $T_{\mu \nu}$}.      Since the bulk in  Gauge/Gravity setups always includes spin-2 fields, and is thought to be a  limit of a truly quantum background-independent theory (``M-theory''), it is not surprising holography should arise and, with the additional constraints, lead to an equation of the type of Eq. \ref{constraint3}.   Also, if Eq. \ref{zdef} applies, 
renormalization of the bulk and the boundary is conducted on an equal basis, since there is one partition function.  

It remains to be seen to what extent this is compatible with the holographic renormalization program \cite{holoreno1,holoreno2}.  However, Equation Eq. \ref{sdef} and footnote 5 make it natural that the limit of classical gravity corresponds to the planar limit on the horizon.     Classical gravity, corresponds to the limit where fluctuations on the horizon vanish (the bulk and the horizon metrics correspond to the classical expectation value with overwhelming proability).  In addition to $S_{int} \rightarrow 0$, this requires that field fluctuations of the horizon theoery are negligible when the total $S_{\partial \Sigma}$ is calculated (indeed, dimensional analysis of Eq. \ref{sdef} makes it natural that $S_{int}$ is related to quantum corrections to $S_{\partial \Sigma}$, $\ave{S_{int}} \sim  \ave{(\Delta S_{\partial \Sigma})^2}$ ).   The planar limit, where there are $\order{N^2}\gg 1$ fields $\phi_i$ with $\left[ \phi_i,\phi_j\right]\sim \delta_{i j}$, and
\begin{equation}
\label{heisen}
 S_{\partial \Sigma}^{classical} \sim N^2 \gg \order{1} \eqcomma \ave{(\Delta S_{\partial \Sigma})^2}_{quantum} \sim \sum_i \left| \left[\phi_i,\phi_j  \right]\right|^2 \sim N \ll S_{\partial \Sigma}^{classical}
\end{equation}
Eq. \ref{heisen} makes these requirements satisfied in a natural way, since $T^{ij}_{\partial \Sigma}$ in Eq. \ref{stresse} will be unaffected by any fluctuations of $\phi_i$.

We should note that, while at the moment we cannot tell whether the constraints described here can be fully satisfied by {\em any} theory, some obvious counter-examples apply:  Free field theories (fields that only interact gravitationally) are trivially excluded, since such a theory does not have a Bremstrahlung process to balance interactions with the Unruh horizon.
More subtly, consider the infalling observer calculation described in section VII of \cite{stojk}.     The authors argue that an infalling observer's detection of Hawking radiation is suppressed by the finite size of the locally available detector with respect to the characteristic
frequency of Hawking's radiation.      Provided the equivalence principle holds semiclassically, this conclusion is inevitable, since otherwise the infalling observer would have a local measurement at their disposal which would tell them they are in fact infalling rather than in an inertial frame.
However, a tacit assumption made in this argument:  That the particle absorption cross-section of such a detector vanishes in the infrared limit.    This assumption is {\em not} obviously related to the equivalence principle, and does {\em not} hold for general field theories:  Obvious counter examples are conformally invariant theories (such as $\mathcal{N}=4$ SYM \cite{holo}) or alternatively massless $\phi^4$ theory.    In a world where such theories were coupled to gravity, tree level quantum effects would violate the equivalence principle and the constraints of Eq. \ref{constraint3} would not hold \footnote{The Corresponding Unruh effect manifestation of this reasoning is the radiation of zero energy photons in the comoving frame by a system which in an inertial frame is undergoing Bremstrahlung \cite{higuchi1,higuchi2}: These photons generally would generally break the identification of Bremstrahlung with the comoving Unruh frame, but are undetectable by any locally comoving detector {\em provided} the interaction cross-section vanishes in the infrared limit }
However, this assumption {\em does} hold true for all theories which seem to be physically relevant, from QED (where it is enforced by Wards identities \cite{qft}) to Yang-Mills and the Higgs sector (where it is enforced by the existence of a mass gap in the infrared regime).     Thus, we can be hopeful that the existence of a meaningful quantum equivalence principle has a role in selecting physically relevant theories.   The generalized second law of thermodynamics (see footnote (3) and references therein) and energy conditions could provide further constraints. 
\section{Discussion and conclusions}
In this work, we have {\em not} outlined a theory of quantum gravity.   The geometry that defines $g_{\Sigma}$ and $g_{\partial \Sigma}$ is classical, with quantum theories living on it.  
In the semiclassical limit the bulk metric's Einstein Tensor $G_{\mu \nu}$ is related to the bulk stress-energy tensor $T_\Sigma^{\mu \nu}$ via Einstein's equation
\begin{equation}
\label{einstein}
G^{\mu \nu} \simeq \ave{G^{\mu \nu} }= \ave{R^{\mu \nu} - \frac{1}{2} g^{\mu \nu}_\Sigma R} = 8 \pi l_p^2  \ave{T^{\mu \nu}_\Sigma}
\end{equation}
which has long been understood as equivalent to an equation for horizon thermodynamics's \cite{jacobson}.    This evolution is also the semiclassical leading term of a Lindblad-type equation \cite{lindblad}, with the super-horizon degrees of freedom traced over instantaneously, consistently with the fact that to leading order in $l_p $, the horizon is determined ``instantaneously'' (in the detectors comoving frame) from the Rayachaduri equation \cite{ray} and $\ave{\hat{T}_{\mu \nu}}$ of the bulk fields.

Rather, we have written down a recipe for constructing an EFT respecting covariance with respect to non-inertial transformations, and hence have paved the way to incorporating the equivalence principle in an effective theory expansion.   Equation \ref{zdef} is classical in gravity and leading order in $l_p$, but it is in arbitrary order in $\hbar$ and EFT coupling constants.  In a regime where a semiclassical EFT is a good approximation, Eq. \ref{constraint1} defines the constraints necessary for a leading term in a series where the EP is satisfied term-by-term. 
That said, Eq. \ref{einstein} and Eq \ref{zdef} allow us to conjecture higher terms in the background independent EFT
\begin{equation}
\label{higherterms} \ave{G_{\mu \nu}^n} \sim l_p^{2n} \frac{\delta^n}{\delta g_{\mu \nu}^n} \ln Z
\end{equation}
However, the neglected $S_{int}$ terms in Eq.\ref{sdef}, of the same order as those in Eq. \ref{higherterms}, are as yet undetermined.  Physically they address the {\em quantum process} of crossing the horizon, something that happens instantaneously up to $\order{ l_p}$.   Such terms, naively related to the fluctuation terms in Eq. \ref{heisen}, might however be crucial to understand the quantum information aspects \cite{gteft} of quantum gravity.
 
We close with a few considerations on possible phenomenological consequences of these ideas.    
Gedankenexperiments such as \cite{eppley}, used to illustrate the contradiction between the quantum partition function in Eqs. \ref{zdef},\ref{constraint1} and the classical treatment of the geometry in Eq. \ref{einstein}, as well as being experimentally unfeasible, are formulated in a regime where such EFTs fail and a transplanckian theory is necessary. 
Hence, fundamentally as well as practically, any hope for testing these ideas should focus on {\em weak} fields and accelerations with respect to the relevant power of $l_p$.

For instance, Gravitational waves in this approach look very different from the usual approaches, centered around ``graviton'' quantization.
 Seen by a detector, a wave in Minkowski space will appear as a horizon that briefly becomes time-like and then goes back to being light-like.   During the ``time-like period'' the gravitational wave will {\em decay} in the rest-frame of the detector.   It is reasonable to assume the gravitational wave dissipates into very soft (with respect to the wave frequency $w$) photons\footnote{Indeed the scenario discussed here could conceivably be ruled out if the observation of cosmological B-modes, claimed by the BICEPS collaboration \cite{biceps2} and put into doubt by Planck \cite{planck}, was confirmed.  The origin of primordial gravitational waves in the inflationary paradigm is, at leading order, due to $\left[ h_{\mu \nu},\Pi_{\mu \nu} \right] \ne 0$ \cite{krauss,anupam}. Other mechanisms, where primordial waves are generated by reaction with matter, however, also exist, through suppressed by powers of $l_p^2$ }, analogously to black hole evaporation and the dissipation of sound in a viscous medium\cite{verlinde}, on a time-scale of $\sim l_p^2 w$.    This effect cannot be captured by perturbation expansions, since the perturbative decay amplitude of a graviton to {\em any} number of photons is zero \cite{genanp}, but can be readily understood as the kind of semiclassical non-perturbative effect background independence requires for arbitrary weak perturbations. This implies that gravitational waves, of arbitrarily small amplitudes, must be thought of as collective ``semiclassical'' excitations.  More quantitative calculations in this direction can be done by the linearized horizon formalism described in \cite{bianchipaper}.

More generally, the strong equivalence principle can be directly experimentally probed to a level necessary to test if it applies beyond linearized GR.
In addition to the lunar ranging experiments mentioned earlier \cite{nor1,nor2}
QED setups in an accelerated frame could lead to regimes where horizon effects appear.  For example, the $\sim \hbar/m$ interference detected in \cite{colella} should appear {\em exactly} in an accelerating frame, and disappear, up to tidal forces, in a freely falling frame (in orbit, for example).  While this is obvious in the first quantization limit, since wavenumber shift$\sim$ momentum change $\sim$ force $\times$ time, in the field theory limit, where back-reaction with the Unruh effect is non-negligible, the applicability of the equivalence principle could yield quantifiable predictions.   Formalisms such as \cite{soph} for QFTs and \cite{das} for quantum mechanics can be used to calculate observables under the assumption that full unitarity is maintained in the presence of gravitational fields.
The general reasoning contained in this paper will lead to an expectation that decoherence will happen at a rate comparable to the ratio of the characteristic size of the system (the minimum of the wave-number of the system's constituents and the system's charge radius) to the distance from the horizon in the systems comoving frame.    This is natural scale, defining relativistic "locality" in a quantum setting.    Given the recent work with ultra-powerful lasers \cite{jan}, perhaps quantitative tests of this type will become possible in the coming decades). 

Finally, it should be noted that a cosmological formulation of the conjecture presented here could provide a foundation to the approach in \cite{kaloper}, given the analogy of the action in \cite{kaloper} with Eq. \ref{zdef}.  Furthermore, a cosmological formulation with {\em conserved charges} in addition to $T^{\mu \nu}$, and hence a bulk-boundary {\em chemical potential} augmenting temperature in Eq. \ref{zdef} will naturally lead to a net conserved charge asymmetry in the bulk, in a manner similar to \cite{dharam}.   It remains to be seen to what extent these can furnish solutions to, respectively, the cosmological constant and the baryogenesis problems, but it is a possible approach to develop a phenomenological extension to the ideas espoused in this work.

 The author acknowledges support from FAPESP proc. 2014/13120-7 and CNPQ bolsa 
de produtividade 301996/2014-8.
We wish to thank Sabine Hossenfelder for constructive suggestions and interesting discussions.


\begin{thebibliography}{10}

\bibitem{ep} Einstein, Albert ``The Meaning of Relativity'' Routledge (2003)

\bibitem{qft}
For demonstrations of how one needs Quantum field theory to 
reconcile relativity, causality and quantum mechanics, as well as an 
exploration on the constraints on possible QFTs, I recommend the 
first two chapters of Peskin and Schroeder's book, {\em An Introduction To Quantum Field Theory},  the first three chapters of {\em The Quantum Theory of fields I} by Steven Weinberg or {\em Quantum Field Theory} by Mark Srednicki

\bibitem{weinberg3}  S. Weinberg, ``the Quantum theory of fields'', volume III, Cambridge University Press 1996

\bibitem{blackinfo} 
  S.~D.~Mathur,
  %``The Information paradox: A Pedagogical introduction,''
  Class.\ Quant.\ Grav.\  {\bf 26}, 224001 (2009)
  [arXiv:0909.1038 [hep-th]].
  %%CITATION = ARXIV:0909.1038;%%
  %162 citations counted in INSPIRE as of 09 Jan 2015

%\cite{Unruh:1976db}
\bibitem{unruh}
  W.~G.~Unruh,
  %``Notes on black hole evaporation,''
  Phys.\ Rev.\  D {\bf 14}, 870 (1976).
  %%CITATION = PHRVA,D14,870;%%

\bibitem{hawking} 
  S.~W.~Hawking,
  %``Black Holes and Thermodynamics,''
  Phys.\ Rev.\ D {\bf 13}, 191 (1976).
  %%CITATION = PHRVA,D13,191;%%
  %540 citations counted in INSPIRE as of 09 Jan 2014


%\cite{Fulling:1972md}
\bibitem{fulling}
  S.~A.~Fulling,
  %``Nonuniqueness of canonical field quantization in Riemannian space-time,''
  Phys.\ Rev.\  D {\bf 7}, 2850 (1973).
  %%CITATION = PHRVA,D7,2850;%%

\bibitem{grove} 
  P.~G.~Grove,
  %``On an Inertial Observer's Interpretation of the Detection of Radiation by Linearly Accelerated Particle Detectors,''
  Class.\ Quant.\ Grav.\  {\bf 3}, 801 (1986).
  %%CITATION = CQGRD,3,801;%%
  %52 citations counted in INSPIRE as of 09 Jan 2014


\bibitem{haag} 
  R.~Haag,
  %``On quantum field theories,''
  Kong.\ Dan.\ Vid.\ Sel.\ Mat.\ Fys.\ Med.\  {\bf 29N12}, 1 (1955)
  [Z.\ Phys.\  {\bf 141}, 217 (1955)]
  [Phil.\ Mag.\  {\bf 46}, 376 (1955)].
  %%CITATION = KDVSA,29N12,1;%%
  %25 citations counted in INSPIRE as of 28 Dec 2014

\bibitem{biso} 
  J.~J.~Bisognano and E.~H.~Wichmann,
  %``On the Duality Condition for Quantum Fields,''
  J.\ Math.\ Phys.\  {\bf 17}, 303 (1976).
  %%CITATION = JMAPA,17,303;%%
  %188 citations counted in INSPIRE as of 20 Jan 2015

\bibitem{sewell} 
  G.~L.~Sewell,
  %``Quantum fields on manifolds: PCT and gravitationally induced thermal states,''
  Annals Phys.\  {\bf 141}, 201 (1982).
  %%CITATION = APNYA,141,201;%%
  %123 citations counted in INSPIRE as of 20 Jan 2015

%\cite{Higuchi:1992we}
\bibitem{higuchi1}
  A.~Higuchi, G.~E.~A.~Matsas and D.~Sudarsky,
  %``Bremsstrahlung and zero energy Rindler photons,''
  Phys.\ Rev.\ D {\bf 45}, 3308 (1992).
  %%CITATION = PHRVA,D45,3308;%%
  %34 citations counted in INSPIRE as of 07 Dec 2014

%\cite{Higuchi:1992td}
\bibitem{higuchi2}
  A.~Higuchi, G.~E.~A.~Matsas and D.~Sudarsky,
  %``Bremsstrahlung and Fulling-Davies-Unruh thermal bath,''
  Phys.\ Rev.\ D {\bf 46}, 3450 (1992).
  %%CITATION = PHRVA,D46,3450;%%
  %51 citations counted in INSPIRE as of 07 Dec 2014


%\cite{Pena:2014uia}
\bibitem{sudarsky} 
  I.~Peña and D.~Sudarsky,
  %``On the Possibility of Measuring the Unruh Effect,''
  Found.\ Phys.\  {\bf 44}, no. 6, 689 (2014).
  %%CITATION = FNDPA,44,689;%%


\bibitem{boulware}
  D.~G.~Boulware,
  %``Radiation From a Uniformly Accelerated Charge,''
  Annals Phys.\  {\bf 124}, 169 (1980).
  %%CITATION = APNYA,124,169;%%
  %75 citations counted in INSPIRE as of 07 Dec 2014

\bibitem{saa}
Camila de Almeida, Alberto Saa, Am.J.Phys. 74 (2006) 154-158,
physics/0506049

\bibitem{soph} 
  N.~E.~J.~Bjerrum-Bohr, J.~F.~Donoghue, B.~R.~Holstein, L.~Planté and P.~Vanhove,
  %``Bending of Light in Quantum Gravity,''
  arXiv:1410.7590 [hep-th].
  %%CITATION = ARXIV:1410.7590;%%

%\cite{Das:2013oda}
\bibitem{das} 
  S.~Das,
  %``Quantum Raychaudhuri equation,''
  Phys.\ Rev.\ D {\bf 89}, no. 8, 084068 (2014)
  [arXiv:1311.6539 [gr-qc]].
  %%CITATION = ARXIV:1311.6539;%%
  %3 citations counted in INSPIRE as of 12 Feb 201

%\cite{Hossain:2014fma}
\bibitem{nounruh1} 
  G.~M.~Hossain and G.~Sardar,
  %``Absence of Unruh effect in polymer quantization,''
  arXiv:1411.1935 [gr-qc].
  %%CITATION = ARXIV:1411.1935;%%

\bibitem{nounruh2} 
  P.~Nicolini and M.~Rinaldi,
  %``A Minimal length versus the Unruh effect,''
  Phys.\ Lett.\ B {\bf 695}, 303 (2011)
  [arXiv:0910.2860 [hep-th]].
  %%CITATION = ARXIV:0910.2860;%%
  %32 citations counted in INSPIRE as of 29 Dec 2014




%\cite{Raychaudhuri:1953yv}
\bibitem{ray} 
  A.~Raychaudhuri,
  %``Relativistic cosmology. 1.,''
  Phys.\ Rev.\  {\bf 98}, 1123 (1955).
  %%CITATION = PHRVA,98,1123;%%
  %150 citations counted in INSPIRE as of 25 Oct 2013

%\cite{Jacobson:1995ab}
\bibitem{jacobson} 
  T.~Jacobson,
  %``Thermodynamics of space-time: The Einstein equation of state,''
  Phys.\ Rev.\ Lett.\  {\bf 75}, 1260 (1995)
  [gr-qc/9504004].
  %%CITATION = GR-QC/9504004;%%

\bibitem{verlinde} 
  E.~P.~Verlinde,
  %``On the Origin of Gravity and the Laws of Newton,''
  JHEP {\bf 1104}, 029 (2011)
  [arXiv:1001.0785 [hep-th]].
  %%CITATION = ARXIV:1001.0785;%%
  %442 citations counted in INSPIRE as of 02 Jan 2015

\bibitem{bianchipaper}
% For further details see, e.g. E.~Bianchi and A.~Satz,
  %``Mechanical laws of the Rindler horizon,''
  Phys.\ Rev.\ D {\bf 87}, 124031 (2013)
  [arXiv:1305.4986 [gr-qc]].\\

\bibitem{wald}
R.M. Wald, {\em General Relativity}, Chicago UP (1984)


\bibitem{snewton}
  R.~Penrose,
  %``On gravity's role in quantum state reduction,''
  Gen.\ Rel.\ Grav.\  {\bf 28}, 581 (1996).
  %%CITATION = GRGVA,28,581;%%
  %163 citations counted in INSPIRE as of 02 Jan 2015

\bibitem{tors} 
  S.~Schlamminger, K.-Y.~Choi, T.~A.~Wagner, J.~H.~Gundlach and E.~G.~Adelberger,
  %``Test of the equivalence principle using a rotating torsion balance,''
  Phys.\ Rev.\ Lett.\  {\bf 100}, 041101 (2008)
  [arXiv:0712.0607 [gr-qc]].
  %%CITATION = ARXIV:0712.0607;%%
  %159 citations counted in INSPIRE as of 30 Jan 2015

\bibitem{review} 
  E.~Berti, E.~Barausse, V.~Cardoso, L.~Gualtieri, P.~Pani, U.~Sperhake, L.~C.~Stein and N.~Wex {\it et al.},
  %``Testing General Relativity with Present and Future Astrophysical Observations,''
  arXiv:1501.07274 [gr-qc].
  %%CITATION = ARXIV:1501.07274;%%

\bibitem{nor1}
  K.~Nordtvedt,
  %``Testing relativity with laser ranging to the moon,''
  Phys.\ Rev.\  {\bf 170}, 1186 (1968).
  %%CITATION = PHRVA,170,1186;%%
  %94 citations counted in INSPIRE as of 02 Jan 2015

\bibitem{nor2}
  E.~G.~Adelberger, B.~R.~Heckel, G.~Smith, Y.~Su and H.~E.~Swanson,
  %``Eotvos experiments, lunar ranging and the strong equivalence principle,''
  Nature {\bf 347}, 261 (1990).
  %%CITATION = NATUA,347,261;%%
  %4 citations counted in INSPIRE as of 02 Jan 2015



%\cite{Gibbons:1977mu}
\bibitem{gibbons}
  G.~W.~Gibbons and S.~W.~Hawking,
  %``Cosmological Event Horizons, Thermodynamics, And Particle Creation,''
  Phys.\ Rev.\  D {\bf 15}, 2738 (1977).
  %%CITATION = PHRVA,D15,2738;%%


\bibitem{feynman}
P.Exner, Open Quantum Systems and Feynman Integrals, D.Reidel publishing

\bibitem{zurek} W. H. Zurek, Rev. Mod. Phys. {\bf 75}, 715 (2003)

\bibitem{caldeira} A. O. Caldeira and A. J. Leggett, Physical Review A {\bf 31}, 1059 (1985).

%\cite{Aharony:1999ti}
\bibitem{holo} 
  O.~Aharony, S.~S.~Gubser, J.~M.~Maldacena, H.~Ooguri and Y.~Oz,
  %``Large N field theories, string theory and gravity,''
  Phys.\ Rept.\  {\bf 323}, 183 (2000)
  [hep-th/9905111].
  %%CITATION = HEP-TH/9905111;%%
  %3101 citations counted in INSPIRE as of 09 Jan 2014


%\cite{Bousso:2002ju}
\bibitem{holo1}
  R.~Bousso,
  %``The Holographic principle,''
  Rev.\ Mod.\ Phys.\  {\bf 74}, 825 (2002)
  [hep-th/0203101].
  %%CITATION = HEP-TH/0203101;%%
  %564 citations counted in INSPIRE as of 09 Jan 2014

%\cite{Susskind:1994vu}
\bibitem{holo2}
  L.~Susskind,
  %``The World as a hologram,''
  J.\ Math.\ Phys.\  {\bf 36}, 6377 (1995)
  [hep-th/9409089].
  %%CITATION = HEP-TH/9409089;%%
  %1642 citations counted in INSPIRE as of 09 Jan 2014




%\cite{Padmanabhan:2009vy}
\bibitem{padma} 
  T.~Padmanabhan,
  %``Thermodynamical Aspects of Gravity: New insights,''
  Rept.\ Prog.\ Phys.\  {\bf 73}, 046901 (2010)
  [arXiv:0911.5004 [gr-qc]].
  %%CITATION = ARXIV:0911.5004;%%

%\cite{Padmanabhan:2002jr}
\bibitem{padma2} 
  T.~Padmanabhan,
  %``The Holography of gravity encoded in a relation between entropy, horizon area and action for gravity,''
  Gen.\ Rel.\ Grav.\  {\bf 34}, 2029 (2002)
  [gr-qc/0205090].
  %%CITATION = GR-QC/0205090;%%



\bibitem{rovelli2}
%\cite{Rovelli:2004tv}
  C.~Rovelli,
  %``Quantum gravity,''
%\href{http://www.slac.stanford.edu/spires/find/hep/www?irn=5994683}{SPIRES entry}
{\it  Cambridge, UK: Univ. Pr. (2004) 455 p}

%\cite{Page:1983uc}
\bibitem{wp1} 
  D.~N.~Page and W.~K.~Wootters,
  %``Evolution Without Evolution: Dynamics Described By Stationary Observables,''
  Phys.\ Rev.\ D {\bf 27}, 2885 (1983).
  %%CITATION = PHRVA,D27,2885;%%
  %121 citations counted in INSPIRE as of 24 Oct 2013

\bibitem{wp2}
K.V. Kuchar, in “General Relativity and Gravitation: Proceedings of the 4th Canadian Conference  on General Relativity and Relativistic Astro-
physics” ed. G. Kunstatter, D. Vincent and J. Williams
7
(World Scientific, Singapore 1992).


%\cite{Gambini:2008ke}
\bibitem{wp3} 
  R.~Gambini, R.~A.~Porto, J.~Pullin and S.~Torterolo,
  %``Conditional probabilities with Dirac observables and the problem of time in quantum gravity,''
  Phys.\ Rev.\ D {\bf 79}, 041501 (2009)
  [arXiv:0809.4235 [gr-qc]].
  %%CITATION = ARXIV:0809.4235;%%
  %23 citations counted in INSPIRE as of 24 Oct 2013


%\cite{Page:1993ij}
\bibitem{wp4} 
  D.~N.~Page,
  %``Clock time and entropy,''
  gr-qc/9303020.
  %%CITATION = GR-QC/9303020;%%
  %7 citations counted in INSPIRE as of 24 Oct 2013

%\cite{Ashtekar:2014kba}
\bibitem{ashtekar} 
  A.~Ashtekar, M.~Reuter and C.~Rovelli,
  %``From General Relativity to Quantum Gravity,''
  arXiv:1408.4336 [gr-qc].
  %%CITATION = ARXIV:1408.4336;%%
  %3 citations counted in INSPIRE as of 04 Jan 2015

%\cite{Chapline:2000en}
\bibitem{chapline} 
  G.~Chapline, E.~Hohlfeld, R.~B.~Laughlin and D.~I.~Santiago,
  %``Quantum phase transitions and the breakdown of classical general relativity,''
  Int.\ J.\ Mod.\ Phys.\ A {\bf 18}, 3587 (2003)
  [gr-qc/0012094].
  %%CITATION = GR-QC/0012094;%%
  %94 citations counted in INSPIRE as of 02 Jan 2015

\bibitem{bekenstein} 
  J.~D.~Bekenstein,
  %``A Universal Upper Bound on the Entropy to Energy Ratio for Bounded Systems,''
  Phys.\ Rev.\ D {\bf 23}, 287 (1981).
  %%CITATION = PHRVA,D23,287;%%
  %472 citations counted in INSPIRE as of 07 Mar 2014

\bibitem{casini} 
  H.~Casini,
  %``Relative entropy and the Bekenstein bound,''
  Class.\ Quant.\ Grav.\  {\bf 25}, 205021 (2008)
  [arXiv:0804.2182 [hep-th]].
  %%CITATION = ARXIV:0804.2182;%%
  %9 citations counted in INSPIRE as of 07 Mar 2014

\bibitem{davies1}
 P.~C.~W.~Davies,
  %``Cosmological Horizons And Entropy,''
  Class.\ Quant.\ Grav.\  {\bf 5}, 1349 (1988).
  %%CITATION = CQGRD,5,1349;%%

%\cite{Davis:2003ye}
\bibitem{davies2}
  T.~M.~Davis, P.~C.~W.~Davies and C.~H.~Lineweaver,
  %``Black hole versus cosmological horizon entropy,''
  Class.\ Quant.\ Grav.\  {\bf 20}, 2753 (2003)
  [astro-ph/0305121].
  %%CITATION = ASTRO-PH/0305121;%%

\bibitem{gribov} 
For a review of this approach See for example \\
 L.~S.~Grigorio, M.~S.~Guimaraes, R.~Rougemont and C.~Wotzasek,
  %``Effective confinement theory from Abelian variables in SU(3) gauge theory,''
  Phys.\ Lett.\ B {\bf 710}, 683 (2012)
  [arXiv:1108.2619 [hep-th]]. \\ In Non-Abelian gauge theories, gauge choice is not possible when non-perturbative field fluctuations are taken into account in the parth integral, but ``equivalent'' gauge choices are not connected by continuus deformations.   Such equivalent gauge choices are separated by a horizon in field strength $A^\mu$.   As the authors of the review quoted above show, one can approximate qualitative features of non-perturbative QCD, like confinement, by limiting the functional integral of the fields to modes below this horizon.
While physically Gauge symmetry is very different from general relativity (in Gauge theory ``background independence'' is defined in an internal space, and observables are Gauge-invariant.   In General relativity it is defined in physical space, and observables are Gauge-dependent), mathematically both are based on diffeomorphism invariance (in internal and physical space respectively), hence the equivalent of horizons, in internal space, must appear and have a role in Gauge theory.   The mathematical consequences of both horizons in $A^\mu$ and horizons in spacetime are equivalent: The appearance of boundary terms in the effective Lagrangian.    Physically, these horizons do very different things;  In GR, the horizon breaks unitarity and represents the unobservability of certain degrees of freedom.   In Non-Abelian gauge theory it does not affect unitarity since it separates Gauge-equivalent regions, but it introduces non-perturbative effects into quark and gluon propagators, and might ultimately be responsible for confinement.   While we are not even close to understanding the full consequences of this, the idea that the confinement phase transition and the physics of black holes are deeply analogous is thus natural.

  %%CITATION = ARXIV:1108.2619;%%
  %1 citations counted in INSPIRE as of 26 Mar 2015


%\cite{Kuchar:1982eb}
\bibitem{kuchar} 
  K.~Kuchar,
  %``Conditional Symmetries In Parametrized Field Theories,''
  J.\ Math.\ Phys.\  {\bf 23}, 1647 (1982).
  %%CITATION = JMAPA,23,1647;%%
  %17 citations counted in INSPIRE as of 29 Dec 2014
%

\bibitem{nondeg0} 
  L.~A.~Cabral and V.~O.~Rivelles,
  %``Particles and strings in degenerate metric spaces,''
  Class.\ Quant.\ Grav.\  {\bf 17}, 1577 (2000)
  [hep-th/9910163].
  %%CITATION = HEP-TH/9910163;%%
  %1 citations counted in INSPIRE as of 29 Dec 2014

\bibitem{nondeg01}
J.~Schray, T.~Dray, C.~A.~Manogue, R.~W.~Tucker and C.~Wang,
  %``The Construction of spinor fields on manifolds with smooth degenerate metrics,''
  J.\ Math.\ Phys.\  {\bf 37}, 3882 (1996)
  [gr-qc/9605037].
  %%CITATION = GR-QC/9605037;%%
  %1 citations counted in INSPIRE as of 29 Dec 2014


%\cite{Schwinger:1959zz}
\bibitem{euclidean} 
  J.~Schwinger,
  %``Euclidean Quantum Electrodynamics,''
  Phys.\ Rev.\  {\bf 115}, 721 (1959).\\
See also F.Guerra, math-ph/0510087
  %%CITATION = PHRVA,115,721;%%
  %61 citations counted in INSPIRE as of 28 Dec 2014


\bibitem{nondeg} 
  H.~K.~Kunduri and J.~Lucietti,
  %``Classification of near-horizon geometries of extremal black holes,''
  Living Rev.\ Rel.\  {\bf 16}, 8 (2013)
  [arXiv:1306.2517 [hep-th]].
  %%CITATION = ARXIV:1306.2517;%%
  %32 citations counted in INSPIRE as of 29 Dec 2014

%\cite{Torrieri:2013lwa}
\bibitem{gteft} 
  G.~Torrieri,
  %``Multi-particle correlations, many particle systems, and entropy in effective field theories,''
  arXiv:1306.5719 [hep-th].
  %%CITATION = ARXIV:1306.5719;%%
  %2 citations counted in INSPIRE as of 30 Dec 2014

\bibitem{gfluid} 
  G.~Torrieri,
  %``Viscosity of An Ideal Relativistic Quantum Fluid: A Perturbative study,''
  Phys.\ Rev.\ D {\bf 85}, 065006 (2012)
  [arXiv:1112.4086 [hep-th]].
  %%CITATION = ARXIV:1112.4086;%%
  %7 citations counted in INSPIRE as of 30 Dec 2014

\bibitem{xiao} 
For a recent review of these attempts, see  X.~Xiao,
  %``Holographic Representation of Local Operators In De Sitter Space,''
  Phys.\ Rev.\ D {\bf 90}, 024061 (2014)
  [arXiv:1402.7080 [hep-th]]\\
or\\
  D.~Sarkar,
  %``(A)dS holography with a cutoff,''
  Phys.\ Rev.\ D {\bf 90}, no. 8, 086005 (2014)
  [arXiv:1408.0415 [hep-th]].
  %%CITATION = ARXIV:1408.0415;%%
  %1 citations counted in INSPIRE as of 23 mar 2015
  %%CITATION = ARXIV:1402.7080;%%
  %2 citations counted in INSPIRE as of 28 Dec 2014


\bibitem{holoreno1} 
  K.~Skenderis,
  %``Lecture notes on holographic renormalization,''
  Class.\ Quant.\ Grav.\  {\bf 19}, 5849 (2002)
  [hep-th/0209067].
  %%CITATION = HEP-TH/0209067;%%
  %459 citations counted in INSPIRE as of 14 Feb 2014


%\cite{Heemskerk:2010hk}
\bibitem{holoreno2} 
  I.~Heemskerk and J.~Polchinski,
  %``Holographic and Wilsonian Renormalization Groups,''
  JHEP {\bf 1106}, 031 (2011)
  [arXiv:1010.1264 [hep-th]].
  %%CITATION = ARXIV:1010.1264;%%
  %117 citations counted in INSPIRE as of 14 Feb 2014



%\cite{Vachaspati:2006ki}
\bibitem{stojk} 
  T.~Vachaspati, D.~Stojkovic and L.~M.~Krauss,
  %``Observation of incipient black holes and the information loss problem,''
  Phys.\ Rev.\ D {\bf 76}, 024005 (2007)
  [gr-qc/0609024].
  %%CITATION = GR-QC/0609024;%%
  %86 citations counted in INSPIRE as of 12 Dec 2014

\bibitem{lindblad} 
For an illustrative example see\\
  T.~Banks, L.~Susskind and M.~E.~Peskin,
  %``Difficulties for the Evolution of Pure States Into Mixed States,''
  Nucl.\ Phys.\ B {\bf 244}, 125 (1984).\\
Note that if this work is correct, energy in the bulk would be conserved only on
average, confirming this paper's derivation
  %%CITATION = NUPHA,B244,125;%%
  %223 citations counted in INSPIRE as of 30 Dec 2014


\bibitem{eppley} K. Eppley and E. Hannah. “The Necessity of Quantizing the Gravitational Field.” Foundations of Physics, {\bf 7}:51–65, 1977.


%\cite{Colella:1975dq}
\bibitem{colella} 
  R.~Colella, A.~W.~Overhauser and S.~A.~Werner,
  %``Observation of gravitationally induced quantum interference,''
  Phys.\ Rev.\ Lett.\  {\bf 34}, 1472 (1975).\\
See also JJ Sakurai, {\em ``Modern Quantum Mechanics''}
  %%CITATION = PRLTA,34,1472;%%
  %309 citations counted in INSPIRE as of 26 Jan 2014

\bibitem{biceps2} 
  P.~A.~R.~Ade {\it et al.}  [BICEP2 Collaboration],
  %``BICEP2 I: Detection Of B-mode Polarization at Degree Angular Scales,''
  arXiv:1403.3985 [astro-ph.CO].
  %%CITATION = ARXIV:1403.3985;%%
  %239 citations counted in INSPIRE as of 29 Apr 2014

\bibitem{planck} 
  R.~Adam {\it et al.}  [Planck Collaboration],
  %``Planck intermediate results. XXX. The angular power spectrum of polarized dust emission at intermediate and high Galactic latitudes,''
  arXiv:1409.5738 [astro-ph.CO].
  %%CITATION = ARXIV:1409.5738;%%
  %137 citations counted in INSPIRE as of 15 Feb 2015

\bibitem{krauss} 
  L.~M.~Krauss and F.~Wilczek,
  %``Using Cosmology to Establish the Quantization of Gravity,''
  Phys.\ Rev.\ D {\bf 89}, 047501 (2014)
  [arXiv:1309.5343 [hep-th]].
  %%CITATION = ARXIV:1309.5343;%%
  %10 citations counted in INSPIRE as of 29 Apr 2014



\bibitem{anupam} 
  A.~Ashoorioon, P.~S.~B.~Dev and A.~Mazumdar,
  %``Implications of purely classical gravity for inflationary tensor modes,''
  arXiv:1211.4678 [hep-th].
  %%CITATION = ARXIV:1211.4678;%%
  %11 citations counted in INSPIRE as of 29 Apr 2014

\bibitem{genanp} 
  G.~Fiore and G.~Modanese,
  %``General properties of the decay amplitudes for massless particles,''
  Nucl.\ Phys.\ B {\bf 477}, 623 (1996)
  [hep-th/9508018].
  %%CITATION = HEP-TH/9508018;%%
  %5 citations counted in INSPIRE as of 23 Sep 2014

\bibitem{jan} 
  B.~M.~Hegelich, G.~Mourou and J.~Rafelski,
  %``Probing the quantum vacuum with ultra intense laser pulses,''
  Eur.\ Phys.\ J.\ ST {\bf 223}, no. 6, 1093 (2014)
  [arXiv:1412.8234 [physics.optics]].
  %%CITATION = ARXIV:1412.8234;%%
  %2 citations counted in INSPIRE as of 02 Jan 2015

\bibitem{kaloper} 
  N.~Kaloper and A.~Padilla,
  %``Sequestering the Standard Model Vacuum Energy,''
  Phys.\ Rev.\ Lett.\  {\bf 112}, 091304 (2014)
  arXiv:1309.6562 [hep-th].
  %%CITATION = ARXIV:1309.6562;%%
  %3 citations counted in INSPIRE as of 07 Mar 2014

\bibitem{dharam}
  D.~V.~Ahluwalia, M.~Kirchbach and ,
  %``Primordial space-time foam as an origin of cosmological matter antimatter asymmetry,''
  Int.\ J.\ Mod.\ Phys.\ D {\bf 10}, 811 (2001)
  [astro-ph/0107246].
  %%CITATION = ASTRO-PH/0107246;%%
  %17 citations counted in INSPIRE as of 08 Oct 2014

\end{thebibliography}
\end{document}